\documentclass[twocolumn,aps]{revtex4}
\usepackage{graphicx}
\usepackage{amsmath}
\begin{document}
\title{Theoretical and numerical investigation of the shock formation of dust ion acoustic waves\footnote{Proceedings of the \textit{International Conference on
Plasma Physics - ICPP 2004}, Nice (France), 25 - 29 Oct. 2004;
contribution P1-049; Electronic proceedings available online at:
\texttt{http://hal.ccsd.cnrs.fr/ccsd-00001947/en/}  .}}
\author{B. Eliasson and P. K. Shukla}
\affiliation{Institut f\" ur Theoretische Physik IV, Fakult\"at f\"ur
Physik und Astronomie, Ruhr--Universit\"at Bochum, D--44780 Bochum, Germany}
\date{Received 12 October 2004}
\begin{abstract}
We present a theoretical and numerical study of the 
self-steepening and shock formation of large-amplitude 
dust ion-acoustic waves (DIAWs) in
dusty plasmas. We compare the
non-dispersive two fluid model, which predicts the formation of
large amplitude compressive and rarefactive dust ion-acoustic
(DIA) shocks, with Vlasov/fluid simulations where ions are
treated kinetically while a Boltzmann distribution is assumed
for the electrons.
\end{abstract}

\maketitle

Shukla and Silin \cite{r1} predicted the existence of small
amplitude dust ion-acoustic waves (DIAWs) in an unmagnetized dusty
plasma. In the DIAWs, the restoring force comes from the pressure
of inertialess electrons, while the ion mass provides the inertia
to support the waves. On the timescale of the DIAWs, charged dust
grains remain immobile, and they affect the overall
quasi-neutrality of the plasma. When the dust grains are charged
negatively, one has the depletion of the electrons in the
background plasma. Subsequently, the phase speed [$\omega/k
=(n_{i0}/n_{e0}+3T_i/T_e)^{1/2}C_s$, where $n_{i0}$ $(n_{e0})$ is
the unperturbed ion (electron) number density, $T_i$ $(T_e)$ is
the ion (electron) temperature, $C_s =(T_e/m_i)^{1/2}$ is the ion
acoustic speed, and $m_i$ is the ion mass] of the DIAWs becomes
larger than the usual ion-acoustic speed in an electron-ion plasma
without negatively charged dust, since $n_{i0}
> n_{e0}$. When $n_{i0} \gg n_{e0}$ or $T_i \gg T_e$, small-amplitude DIAWs do not
suffer Landau damping in a plasma, since the increased phase speed
is much larger than the ion thermal speed $(T_i/m_i)^{1/2}$.
Small-amplitude DIAWs have been observed in laboratory experiments
\cite{r2}, and the observed phase speed is in an excellent
agreement with the theoretical prediction of Ref. \cite{r1}.

Recently, laboratory  experiments \cite{r3,r4,r5,r6} have been
conducted to study the formation of dust ion-acoustic (DIA) shocks
in dusty plasmas. Dust ion acoustic compressional pulses have been
observed to steepen as they travel through a plasma containing
negatively charged dust grains. Theoretical models \cite{r7,r8}
have been proposed to explain the formation of small amplitude DIA
shocks in terms of the Korteweg-de Vries-Burgers equation, in
which the dissipative terms comes from the dust charge
perturbations \cite{r9}. Popel {\it et al.} \cite{r10} have
included sources and sinks in the ion continuity equation, linear
ion pressure gradients in the nonlinear ion momentum equation with
a model collision term, as well as the dust grain charging
equation to study the formation DIA shock-wave structures.

In this Brief Communication, we present analytical and numerical
studies of large amplitude DIA shock waves in an unmagnetized
dusty plasma \cite{PoP}. We use fully nonlinear continuity and momentum
equations for the warm ion fluid, as well as Boltzmann distributed
electrons and the quasi-neutrality condition to examine the
spatio-temporal evolution of large amplitude dust ion-acoustic
pulses. We find simple-wave solutions of our fully nonlinear two
fluid model, and compare them with those deduced from the
time-dependent Vlasov simulations which uses initial conditions
corresponding to the ones obtained from our theoretical model.

We consider an unmagnetized dusty plasma whose constituents are
singly charged positive ions, electrons and charged dust grains.
Thus, at equilibrium, we have $n_{i0} =n_{e0} - \epsilon Z_d
n_{d0}$, where $\epsilon$ equals $-1$ $(+1)$ for negatively
(positively) charged dust grains, $Z_d$ is the number of
elementary charges residing on the dust grain, and $n_{d0}$ is the
equilibrium dust number density. On the timescale of our interest,
the dust grains are assumed to be immobile. The dynamics of low
phase speed (in comparison with the electron thermal speed)
nonlinear, dust ion-acoustic waves is governed by a Boltzmann
distribution for the electrons

\begin{equation}
n_e = n_{e0}\exp\left(\frac{e\phi}{T_e}\right),
\end{equation}
and the continuity and momentum equations for the ions

\begin{equation}
  \frac{\partial n_i}{\partial t}+\frac{\partial (n_i v_i)}{\partial x}=0,
\end{equation}
and

\begin{equation}
  \frac{\partial v_i}{\partial t}+v_i\frac{\partial v_i}{\partial x}=
  -\frac{e}{m_i}\frac{\partial \phi}{\partial x}-\frac{3T_i n_i}
  {m_i n_{i0}^2}\frac{\partial n_i}{\partial x},
\end{equation}
where $n_e$ $(n_i)$ is the total electron (ion) number density,
$e$ is the magnitude of the electron charge, $\phi$ is the wave
potential, and $v_i$ is the ion fluid velocity. The system is
closed by means of Poisson's  equation

\begin{equation}
  \frac{\partial^2\phi}{\partial x^2}=4\pi e(n_e-n_i-\epsilon Z_d n_{d0}).
\end{equation}

In the following, we consider non-dispersive DIAWs, and use the
quasi-neutrality condition $n_e = n_i +\epsilon Z_d n_{d0}$
instead of Eq. (4), together with the normalized variables
$N=n_i/n_{i0}$, $u=v_i/C_s$, $\varphi=e\phi/T_e$, $z=r_D^{-1} x$
and $\tau=\omega_{pi}t$, where $\omega_{pi}=(4\pi
n_{i0}e^2/m_i)^{1/2}$ is the ion plasma frequency and
$r_D=C_s/\omega_{pi}$ is the electron Debye radius. Thus, the
system of equations (1)-(3) can be rewritten as

\begin{equation}
  \frac{\partial u}{\partial \tau}+u\frac{\partial u}{\partial z}+
  \left(\frac{1}{N+\alpha-1}+3\eta N\right)\frac{\partial N}{\partial z}=0,
\end{equation}
and
\begin{equation}
  \frac{\partial N}{\partial \tau}+ N\frac{\partial u}{\partial z}+
  u\frac{\partial N}{\partial z}=0,
\end{equation}
where $\alpha=n_{e0}/n_{i0}$ and $\eta=T_{i0}/T_e$. In obtaining
Eq. (5), we have used $\varphi=\mathrm{ln}[(N+\alpha-1)/\alpha]$
which follows from Eq. (1) and the quasineutrality condition.

In order to study the nonlinear evolution of large amplitude
DIAWs, we seek simple wave solutions \cite{r11} of Eqs. (5) and
(6). For this purpose, we rewrite them in the matrix form as

\begin{equation}
\frac{\partial}{\partial \tau}
\left[
\begin{matrix}
u
\\
N
\end{matrix}
\right]
+
\left[
\begin{matrix}
u & \frac{1}{N+\alpha-1}+3\eta N
\\
N & u
\end{matrix}
\right]
\frac{\partial}{\partial z}
\left[
\begin{matrix}
u
\\
N
\end{matrix}
\right]
=
\left[
\begin{matrix}
0
\\
0
\end{matrix}
\right].
\end{equation}
Here, the nonlinear wave speeds are given by the eigenvalues

\begin{equation}
  \lambda_{\pm}=u\pm N^{1/2}\left(\frac{1}{N+\alpha-1}+3\eta N \right)^{1/2}
\end{equation}
of the square matrix multiplying the second term in Eq. (7).
The square matrix in Eq. (7), which we denote $A$, can be diagonalized by
a diagonalizing matrix $C$ whose columns are the eigenvectors of $A$, so that

\begin{equation}
  C^{-1}AC=\Lambda\equiv \left[
  \begin{matrix}
  \lambda_+ & 0
  \\
  0 & \lambda_-
  \end{matrix}
  \right],
\end{equation}
where

\begin{equation}
  C=\left[
  \begin{matrix}
  1 & 1
  \\
  \left(\frac{1}{N(N+\alpha-1)}+3\eta \right)^{-1/2}&
  -\left(\frac{1}{N(N+\alpha-1)}+3\eta \right)^{-1/2}
  \end{matrix}
  \right],
\end{equation}
and

\begin{equation}
  C^{-1}=\left[
  \begin{matrix}
  \frac{1}{2} & \frac{1}{2}\left(\frac{1}{N(N+\alpha-1)}+3\eta \right)^{1/2}
  \\
  \frac{1}{2} &-\frac{1}{2}\left(\frac{1}{N(N+\alpha-1)}+3\eta \right)^{1/2}
  \end{matrix}
  \right].
\end{equation}

Multiplying Eq. (7) by $C^{-1}$ from the left gives the diagonalized system of equations

\begin{equation}
  \frac{\partial \psi_{+}}{\partial \tau}+\lambda_{+}\frac{\partial \psi_{+}}{\partial z}=0,
\end{equation}
and

\begin{equation}
  \frac{\partial \psi_{-}}{\partial \tau}+\lambda_{-}\frac{\partial \psi_{-}}{\partial z}=0,
\end{equation}
where the new variables are

\begin{equation}
  \psi_{\pm}=\frac{u}{2}\pm \frac{1}{2}
  \int_1^N \left(\frac{1}{N'(N'+\alpha-1)}+3\eta \right)^{1/2}\, dN'.
\end{equation}

Equations (12) and (13) describe the DIAWs propagating in the
positive and negative $z$ directions, respectively. Setting
$\psi_{-}$ to zero, we have

\begin{equation}
  u(N)=\int_1^N [1/N'(N'+\alpha-1)+3\eta ]^{1/2}\, dN',
\end{equation}
 which inserted into Eq. (12) gives

\begin{equation}
  \lambda_{+}(N)=\int_1^N \left(\frac{1}{N'(N'+\alpha-1)}+3\eta \right)^{1/2}\, dN'
  +N^{1/2}\left(\frac{1}{N+\alpha-1}+3\eta N \right)^{1/2}.
\end{equation}
Since $\psi_{+}$ can be written as a function of $N$, Eq. (12) holds also for $N$,
i.e.

\begin{equation}
  \frac{\partial N}{\partial \tau}+\lambda_{+}(N)\frac{\partial N}{\partial z}=0,
\end{equation}
which, as long as $N$ is continuous, has the general solution
$N=f_0(\xi)$, where $\xi=x-\lambda_{+}(N)t$ and $f_0$ is the
initial condition for $N$. We have plotted $\lambda_{+}$ as a
function of $N$ in Fig. 1. 
\begin{figure}[floatfix]
\includegraphics[width=\columnwidth]{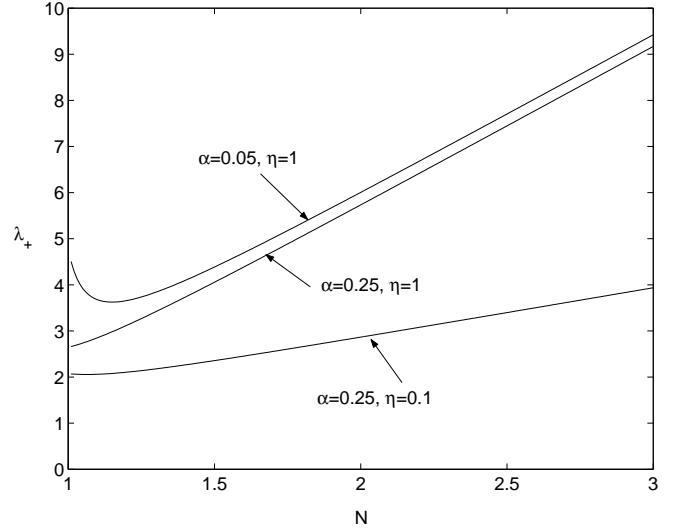}
\caption{
The wave speed $\lambda_{+}$ as a function of $N$ for
different values on $\eta$ and $\alpha$.
}
\end{figure}
Here, we see that $\lambda_{+}$ grows
with increasing $N$ in the two cases with $\alpha=0.25$. In the
case with $\lambda=0.05$, however, the phase speed first decreases
for $N\approx 1$ before increasing with increasing $N$. In the
small-amplitude limit, {\it viz.} $N=1+N_1$, where $|N_1|<<1$, we
have the first-order Taylor expansion $\lambda_{+}=c+\gamma N_1$,
where $c=(1/\alpha+3\eta)^{1/2}$ is the linear acoustic speed and
$\gamma=(3\alpha+12\eta\alpha^2-1)/2\alpha(\alpha+3\eta\alpha^2)^{1/2}$
is the coefficient in front of the nonlinear term. We note that
$\gamma$ is negative for sufficiently small $\alpha$ (in agreement
with the $\alpha=0.05$ case displayed in Fig. 1), and in the cold
ion limit ($\eta=0$) we recover the result that $\alpha<1/3$ leads
to a negative coefficient \cite{r8,r12} in front of the nonlinear
term. The linear acoustic speed increases when $\alpha$ decreases.
Thus, in the presence of negatively charged dust, the phase speed
of the waves may becomes much larger than the ion acoustic speed,
so that the Landau damping of the waves decreases \cite{r1}.

In order to compare the fluid and kinetic theories, we have solved
the coupled Eqs. (5) and (6) numerically and compared the results
with numerical solutions of the Vlasov equation. As an initial
condition for our fluid simulations, we take a large-amplitude
localized density pulse, $N=1.5-0.5\mathrm{sech}[3\sin(2\pi
z/20000)+1.5]$, while the initial condition for the velocity is
obtained from the simple wave solution as $u(N)=\int_1^N
[1/N'(N'+\alpha-1)+3\eta ]^{1/2}\, dN'$. The results are compared
with numerical solutions of the ion Vlasov equation
\begin{equation}
  \frac{\partial f}{\partial \tau}+v\frac{\partial f}{\partial z}
  +\frac{\partial \varphi}{\partial z}\frac{\partial f}{\partial v}=0,
\end{equation}
where $v$ has been normalized by $C_s$ and the ion distribution
function $f$ by $n_{i0}/C_s$. Here, we have also used the
quasineutrality condition and thus
$\varphi=\mathrm{ln}[(N+\alpha-1)/\alpha]$, where
$N=\int_{-\infty}^{\infty}f\,dv$. For the initial condition, we
are using the shifted Maxwellian ion distribution function
\begin{equation}
  f(z,v)=\frac{N(z)}{\sqrt{2\pi\eta T(z)}}\exp\left[
  -\frac{[v-u(z)]^2}{2\eta T(z)}
  \right],
\end{equation}
where we are using the same initial condition for the density $N$
and the velocity $u$ as in the fluid simulations. For the scaled
(by $T_{i0}$) ion temperature $T$, we obtain an initial condition
by combining the ideal gas law $P=N T$, where $P=P_i/P_{i0}$ is
the normalized ion pressure and the adiabatic law $P=N^3$, giving
the initial condition $T=N^2$.

\begin{figure}[floatfix]
\includegraphics[width=\columnwidth]{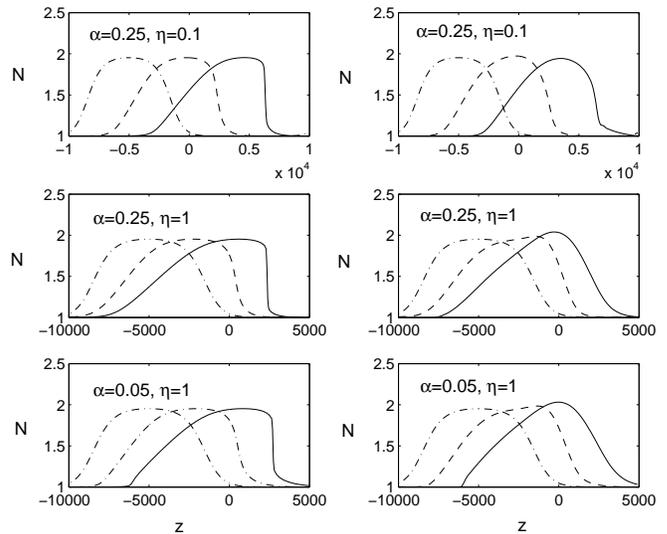}
\caption{
The profile of the ion density $N$ obtained
from numerical solutions of the fluid equations (left panels) and
the Vlasov equation (right panels). In the upper panels, the
profiles are shown at $t=0$ (dash-dotted lines), $t=1700$ (dashed
lines) and $t=3400$ (solid lines), while in the middle and lower
panels the profiles are shown at $t=0$, $t=500$ and $t=1000$.
Parameters are $\eta=0.1$ and $\alpha=0.25$ (upper panels),
$\eta=1$ and $\alpha=0.25$ (middle panels) and $\eta=1$ and
$\alpha=0.05$ (lower panels).}
\end{figure}
In Fig. 2, we present a comparison between the density profiles
obtained from the fluid and Vlasov simulations, at different
times. In the upper panel, the ion-electron temperature ratio
$\eta=0.1$, and the electron-ion density ratio $\alpha=0.25$. We
see that both the fluid (left) and Vlasov (right) solutions
exhibit shocks, where the shock front is distinct in the fluid
solution and more diffuse in the Vlasov solution. The
corresponding ion distribution function is displayed in Fig. 3. 
\begin{figure}[floatfix]
\includegraphics[width=\columnwidth]{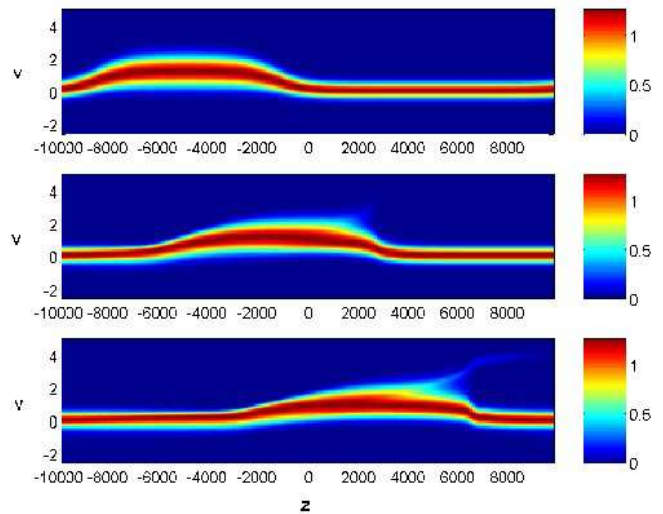}
\caption{
The ion distribution function at $t=0$ (upper
panel), $t=1700$ (middle panel) and $t=3400$ (lower panel) as a
function of $x$ and $v$. Parameters are $\eta=0.1$ and
$\alpha=0.25$.
}
\end{figure}
We
observe that the formation of the shock at $t=3400$ is located at
$z\approx 7000$. It is associated with a ``kink'' in the
distribution function. A population of ions have also been
accelerated by the shock. The middle panels of Fig 2 are for
$\eta=0.1$ and $\alpha=0.25$. Here, the fluid solution exhibits
clear shocks, while the Vlasov simulation shows only a phase of
self-steepening at $t=500$, followed by an expansion of the
diffuse shockfont at $t=1000$. 
\begin{figure}[floatfix]
\includegraphics[width=\columnwidth]{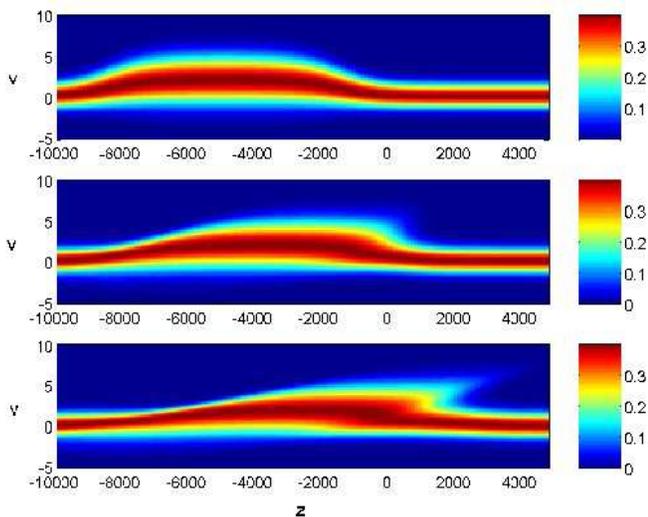}
\caption{
The ion distribution function at $t=0$ (upper
panel), $t=500$ (middle panel) and $t=1000$ (lower panel) as a
function of $x$ and $v$. Parameters are $\eta=1$ and
$\alpha=0.25$.
}
\end{figure}
The ion distribution function in
Fig. 4 shows that the shockfront is strongly Landau damped for
this case. Finally, the bottom panels of Fig. 2 show results for
$\alpha=0.05$ and $\eta=1$. In this case, the fluid solution again
shows a shock in the front end of the pulse, but also the rear end
of the shock steepens, which can be seen at $z\approx -6000$ for
$t=1000$. The steepening of the pulse for low-amplitude density
perturbations in the rear of the pules can be explained by that
the wave speed \emph{decreases} for small-amplitude density
perturbations ($N< 1.1$), as seen in Fig. 1, while it increases
again for large-amplitude density perturbations. The Vlasov
solution again shows a diffusive shock in the front, while it
reproduces the steepening of the density in the rear of the pulse.
\begin{figure}[floatfix]
\includegraphics[width=\columnwidth]{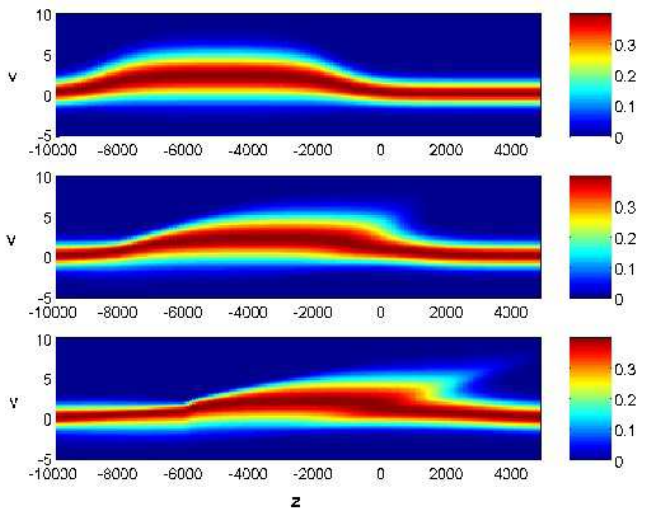}
\caption{
The ion distribution function at $t=0$ (upper
panel), $t=500$ (middle panel) and $t=1000$ (lower panel) as a
function of $x$ and $v$. Parameters are $\eta=1$ and
$\alpha=0.05$.
}
\end{figure}
In the middle panel of Fig. 5, the ion distribution function shows
the self-steepening phase of the shockfront, and the lower panel
shows the diffusion of the shock by shock-accelerated ions. In the
rear end of the pulse, the distribution function forms a ``kink,''
clearly seen at $z\approx -6000$ in the bottom panel.

We have also performed simulations with smaller amplitudes of the
pulses (not shown here) and they exhibit essentially the same
behavior as in the large-amplitude case. It is interesting to note
that it is the strongly heated and shock-accelerated ions in the
pulse that leads to Landau damping by overtaking the pulse. The
heating of the ions is due to adiabatic compression, leading to a
higher thermal speed of the ions inside the pulse than in the
equilibrium plasma. Another effect is that the fluid (mean)
velocity of the ions further accelerates the ions. For Landau
damping to be unimportant, we thus have the condition that the
wave speed must be much larger than the sum of the ion thermal and
fluid velocities. Inserting Eqs. (15) and (16) into the inequality
$\lambda_{+}\gg V_T+u$, where the scaled ion thermal speed
$V_T=(\eta T)^{1/2}\approx\eta^{1/2} N$, we obtain the condition
$N^{1/2}[1/(N+\alpha-1)+3\eta N]^{1/2}\gg(\eta N^2)^{1/2}$, or
$(1/[(N+\alpha-1)N\eta]+3)^{1/2}\gg 1$. This condition is
fulfilled if $\eta\ll 1$ (leading to the sharp shock seen in Fig.
3 and the upper right panel of Fig 2) or/and if the electrons are
evacuated due to the dust so that $\alpha\ll 1$, and at the same
time $N\approx 1$. The latter corresponds to the case where the
sign of the coefficient in front of the low-amplitude nonlinear
term becomes negative, so that there will be a shock in the
\emph{rear} end of the pulse while the front of the shock expands,
in agreement with the observations in Fig. 5 and lower right panel
of Fig 2. The expansion of the shockfront at high dust densities
has also been observed in the experiment \cite{r3}.

To summarize, we have presented the dynamics of fully nonlinear,
nondispersive dust ion acoustic waves in an unmagnetized dusty
plasma. By using the Boltzmann electron distribution as well as
the hydrodynamic equations for the warm ion fluid and
quasi-neutrality condition, we have represented the governing
equations in the form of a master equation whose characteristics
have been found analytically. The fluid equations has been solved
to obtain the density and velocity profiles of the DIA shock
waves, which exhibit the steepening of the waveforms both in the
front and rear depending upon the values of $\alpha$. We have also
compared our theoretical results with those obtained from computer
simulations of the time dependent Vlasov equation. The Vlasov
solution shows a diffuse shock in the front end of the pulse, due
to strong Landau damping, while a sharp shock develops in the rear
end of the pulse, similar to the results from the simulation of
Eqs. (5) and (6).

\acknowledgments

This work was partially supported by the Deutsche Forschungsgemeinschaft (Bonn, Germany)
through the Sonderforschungsbereich 591 and by DOE grant No. DE-FG02-03ER54730.

\end{document}